\documentclass[twocolumn,showpacs,amsmath,amssymb,pre]{revtex4}
\usepackage{graphicx,color,longtable}

\begin{document}
\title{Experimental study of random close packed colloidal particles}
\author{Rei Kurita and Eric R Weeks}
\affiliation{Department of Physics, Emory University, Atlanta, GA 30322 U.S.A}

\date{\today}

\begin{abstract}
A collection of spherical particles can be packed tightly together
into an amorphous packing known as ``random close packing'' (RCP).
This structure is of interest as a model for the arrangement of
molecules in simple liquids and glasses, as well as the arrangement
of particles in sand piles.  We use confocal microscopy to study the
arrangement of colloidal particles in an experimentally realized
RCP state.  We image a large volume containing more than 450,000
particles with a resolution of each particle position to better
than 0.02 particle diameters.  While the arrangement of the
particles satisfies multiple criteria for being random, we also
observe a small fraction (less than 3\%) of tiny crystallites
(4 particles or fewer).  These regions pack slightly better and
are thus associated with locally higher densities.  The structure
factor of our sample at long length scales is non-zero, $S(0) =
0.049 \pm 0.008$, suggesting that there are long wavelength density
fluctuations in our sample.  These may be due to polydispersity
or tiny crystallites.  Our results suggest that experimentally
realizable RCP systems may be different from simulated RCP systems,
in particular, with the presence of these long wavelength density
fluctuations.
\end{abstract}
\maketitle

\section{Introduction}
Dense packings of hard spheres are an important starting point for
the study of simple liquids, metallic glasses, colloids, biological
systems, and granular matter \cite{Bernal,Finney,Scott,Jodrey,
Tobochnik, Lubachevsky, Torquato2005,chaudhuri10}.  Of particular interest
is the densest possible packing that still possesses random
structure, ``random close packing'' (RCP), which is important for
physics and engineering.  For example, the viscosity of dense
particle suspensions diverges when the particles approach the
RCP state \cite{Krieger}.  In a classic experiment, Bernal and
Mason obtained the volume fraction of RCP $\phi_{RCP} \approx
0.637$.  They compressed and shook a rubber balloon which was
full of metal ball bearings for a sufficiently long time to
achieve maximum density \cite{Bernal}.  Scott and Kilgour also
reported $\phi_{RCP} \approx 0.637$ by pouring balls into a
large vibrating container \cite{Scott}.  Their results were
sensitive to the experimental method, for example both the
frequency and amplitude of vibration.  Likewise in computer
simulations, the value of $\phi_{RCP}$ depends on the protocol.
$\phi_{RCP}$ is between 0.642 and 0.648 with a rate dependent
densification algorithm \cite{Jodrey}, 0.68 with a Monte Carlo
methods \cite{Tobochnik} and 0.644 with Lubachevsky-Stillinger
packing algorithm \cite{Lubachevsky,Torquato2005}.  All of these
results are for monodisperse spheres, in other words, spheres with
identical diameters.

The variety of results for $\phi_{RCP}$, in addition to being due to
the method of preparing the RCP state, perhaps also comes from the
poor definition of RCP \cite{Torquato2000,radin08}.  The phrase ``random
close packing" is composed of two terms, ``random" and ``close
packing," which are inherently in conflict with each other.
An ideal \textit{random} state would have no correlation between
particles, but the constraint that particles cannot overlap already
diminishes the randomness of a physical packing.  Furthermore, to
get a \textit{close packing} the most efficient method is to pack
particles into a crystalline array, which is highly non-random
\cite{Hales}.  For example, a random arrangement of spheres
can be made denser if it partially includes dense crystalline
regions, but then it is less random \cite{Davis,Pouliquen}.
In 2000 Torquato and coworkers proposed ``Maximum randomly jammed
(MRJ)" state as a more tight definition of RCP.  MRJ states
are defined as the least locally ordered structures which are
also jammed so that no particles can move \cite{Torquato2000}.
A strictly jammed state should be incompressible and unshearable
\cite{Torquato2003}, while other definitions of jammed states
can involve external forces \cite{Cates} or experimental time
scales \cite{Liu}; the latter can involve questions of glassiness.
Returning to strictly jammed states, one method of quantifying
jamming is by considering the isothermal compressibility $K_T$,
which is determined by the structure factor at wave number $q=0$,
$K_T =  1/\rho (\partial \rho/ \partial p) = S(0)/\rho k_BT$
where $\rho$, $p$, $k_B$ and $T$ are density of the material,
pressure, Boltzmann constant, and temperature, respectively.  Thus,
a strictly jammed state requires $K_T = S(0) = 0$ since this state
should be incompressible.  Indeed, prior simulation works for the
strict jammed state of hard spheres show $S(0) \approx 0$ to within
numerical resolution \cite{Torquato2005,Torquato2003,Silbert}.
The observation $S(q \rightarrow 0) \rightarrow 0$ has been termed
``hyperuniformity,'' in that the density looks increasingly
uniform when considered on longer length scales
\cite{Torquato2005}.

The first physics study of the internal structure of a random closed
packed system that we are aware of is the work of Smith, Foote,
and Busang \cite{smith29}.  In 1929, they studied the packing of
shot and used acid to mark the contacts between spheres, reporting
the contact numbers for 1,562 particles taken from the interior
of a sample with 2,400 particles.  In the 1960's, Bernal first
studied 500 particles taken from the interior of a sample with 5,000
particles \cite{Bernal}, and later 1,000 particles \cite{bernal64}.
In more recent times, 16,000 spheres were studied by Slotterback
{\it et al.} who used an index-matching fluid and laser-sheet
illumination to find the positions \cite{Losert}.  Aste {\it et al.}
used x-ray tomography to study several different granular packings
containing 90,000 particles in an interior region \cite{Aste}.
These experiments provide useful data for testing theories and
studying the properties of RCP packings on large length scales.

In this article, we use a sedimented dense colloidal suspension as
an experimental realization of a RCP material, in the loose sense
of RCP rather than the strict sense of a MRJ state.  We study the
detailed structure of our sample with confocal microscopy, which can
determine the three-dimensional positions of the particle centers
to high accuracy.  By carefully imaging overlapping regions,
we observe a large volume containing over 450,000 particles.  Our
data set is available online \cite{epaps}.  
The data are used to determine which features of our realistic RCP
system are similar to the stricter ideal MRJ packing.  The sample
satisfies several criteria for randomness, for example, having
only a tiny fraction of particles having even short-range order.
However, in contrast to simulated MRJ packings, we find the
isothermal compressibility is not zero, thus suggesting that in
at least this particular experimental realization of a RCP system,
there are differences with simulations.

It is important to note that our colloidal experiment differs
in several particulars from both granular experiments (such
as the early ones with ball bearings \cite{Bernal,Scott})
and simulations.  First, the particles are not all identical;
they have a polydispersity of 5\% in their diameters.  Second,
as the RCP state is formed by sedimentation, the particles have a
chance to diffuse due to Brownian motion.  In some situations this
motion can help particles rearrange into crystalline packings, if
the sample has a volume fraction in the range  $0.49 \le \phi \le
0.58$ \cite{Alder,Pusey}.  While our experimental preparation method
avoids full crystallization, it is plausible that the sample could
be more ordered as a result of subtle rearrangements as particles
sediment toward their final positions.  However, conventionally
such sedimented colloidal samples are thought of as RCP states.
Our primary motivation is to use our sample to discern properties
of the RCP state, and test the applicability of ideas derived
from simulation.

\section{Method}
\subsection{Sample preparation}
We use poly(methyl methacrylate) (PMMA) particles sterically
stabilized with poly-12-hydroxystearic acid \cite{Antl}.
To visualize the particles, they are dyed with rhodamine 6G
\cite{Dinsmore}.  The mean diameter $d$ of our
particles is $d = 2.53$~$\mu$m with an uncertainty 1\%. 
Additionally the particles have a polydispersity of $\sim 5$\%. 
According to prior simulations, the volume fraction
for random close packing $\phi_{RCP}$ is between 0.64 and 0.66
for a 5\% polydisperse system \cite{Farr,schaertl94,hermes10},
which is almost same as $\phi_{RCP}$ for monodisperse spheres.
References \cite{chaudhuri10,Torquato2000,hermes10} point out that the specific
value often depends on the simulation details.

\begin{figure}[htbp]
\begin{center}
\includegraphics[width=6cm]{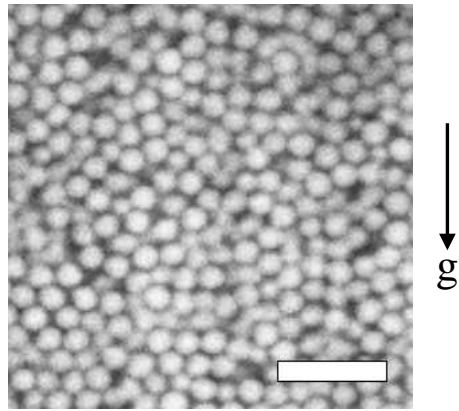}
\end{center}
\caption{Confocal micrograph of the colloidal sediment in ($x$, $y$) plane.
The image was taken 30 $\mu$m inside the
sample. The scale bar represents 10 $\mu$m.
The arrow indicates the direction of gravity during sedimentation.
}
\label{Image}
\end{figure}

We use a fast laser scanning confocal microscope (VT-Eye, Visitech)
which yields clear images deep inside our dense samples.  Despite
the high density, the colloidal particles can be easily discerned as
shown in Fig.~\ref{Image}.  We acquire three-dimensional (3D) scans
of our sample yielding a $62.7 \times 65.4 \times 30$~$\mu$m$^3$
observation volume for each image.  As the sample is jammed,
particles do not move and we can scan slowly to achieve very
clean images:  each 3D scan takes about 30~s.
Within each 3D image, particles are identified within
0.03~$\mu$m in $x$ and $y$, and within 0.05~$\mu$m in $z$
\cite{Dinsmore,Crocker96}.

The PMMA particles are initially suspended in a mixture of 85\%
cyclohexylbromide and 15\% decalin by weight. This mixture closely
matches both the density and refractive index of the particles
\cite{Dinsmore}.  Then, to induce the particles to sediment, we add
a small amount of decalin to slightly decrease the density of the
dispersion fluid. 

We can quantify the importance of sedimentation by computing the
nondimensional Peclet number.  This is the ratio of the time for
a particle to diffuse its own radius $d/2$ to the time for it
to fall a distance $d/2$.  The diffusion time scale is $\tau_D
= d^2/(8D)$, using the diffusion constant $D$, which for our
particles and solvent is $D = 0.1$~$\mu$m$^2$/s.  This gives
us $\tau_D = 6$~s.  The sedimentation time scale is $\tau_S =
d/(2 v_{\rm sed})$.  We observe the height of the sediment as a
function of time in a macroscopic sample of dilute particles, and
find $v_{\rm sed} = 0.035$~$\mu$m/s, giving us $\tau_S = 32$~s.
Thus Pe $\approx 0.2$, suggesting that particles can diffuse long
distances while they sediment; an alternate implication is that
hydrodynamic interactions between particles due to sedimentation
are perhaps less important than diffusion \cite{Segre}.  Prior to
the start of sedimentation, the initial volume fraction is about
0.30 (in the stable liquid phase).  We stir the particles by
ultrasonic wave before sedimentation to avoid the Rayleigh-Taylor
instability \cite{Wysocki}.  
During sedimentation, the sample
passes through the volume fraction range where crystals can be
nucleated, approximately $0.51 < \phi < 0.60$ for our sample with
5\% polydispersity \cite{pusey09,fasolo04}.  We do not observe
crystals in our final sample, and the most likely explanation is
that sedimentation happens faster than nucleation, which is quite
slow for polydisperse samples \cite{pusey09,auer01,schope07}.
For samples with $Pe < 0.1$, we do observe crystallization,
although we have not carefully studied the critical $Pe$ for
which crystallization is suppressed; see Ref.~\cite{hermes10} for
further discussion.

Given that diffusion is faster than sedimentation, the sample
readily equilibrates, at least at low volume fractions as the
sedimentation starts.  Hence,
we believe that our final state is well-defined and
insensitive to the initial state.
During sedimentation, the Stokes drag force acting on the
particles is given by $F = 3 \pi \eta d v$, with viscosity
$\eta=2.18$~mPa$\cdot$s and $v=v_{\rm sed}$.  The buoyant weight
of the particles is given by $W_b = \Delta \rho \pi d^3 g/6$
with $g$ the acceleration due to gravity and $\Delta \rho$ being
the density difference.  Balancing the gravitational force with
the drag force, we can estimate the density difference as $\Delta
\rho = 0.038$~g/cm$^3$.  For reference, the particle density is
1.2340 g/cm$^3$.  Balancing the gravitational energy $W_b h$ with
the thermal energy $k_B T$ lets us determine the scale height $h =
1.8$~$\mu$m (using $k_B$ as Boltzmann's constant).  The small scale
height suggests that in the final sedimented state, there will be
no density stratification except right at the interface between
the dense sediment and the remaining solvent; that interface will
have a thickness $O(h)$.

During the sedimentation process, it takes about 2 days for the
sample to initially sediment to the bottom and form a glassy
state.  However, the sedimentation speed is slow at high $\phi$
\cite{Paddy,Marconi}.  Thus, we wait 90 days to complete the
sedimentation before we put the sample on
the microscope.  We also re-checked the
sample 300 days after the initial sediment, and found the same
results as a 90 day old sample.

We use the convention that the $y$ direction is the axis
corresponding to gravity during the sedimentation process (see
Fig.~\ref{Image}).  
The sample chamber is made from glass slides and
coverslips, sealed with UV-curing epoxy (Norland), with the sample dimensions
being $x=6$~mm, $y=20$~mm, and $z=0.14$~mm.  When we measure the
structure, we lay our sample on the microscope; that is, the optical
$z$ axis is parallel to gravity and the microscope looks into the
thinnest dimension of the sample chamber (for ease of viewing).
In the highly concentrated sample, any subsequent gravity-induced
particle rearrangements are much slower than our measuring time.
In particular, we do not observe any particle flow in the sample,
and the structure does not change at all during measurement.
Near the flat coverslip of our sample chamber, particles layer
against the wall \cite{Ken,nugent07prl}.  To avoid influence of
this, we take our 3D images at about 1 mm above the $y$ axis sample
chamber bottom and at about 15 $\mu$m above the glass slide along
the $z$ axis.  Simulations show that wall effects decay fairly
rapidly ($\sim 4$ diameters = 10 $\mu$m) \cite{Ken}, and in our data
we see no density fluctuations as a function of the distance $z$
away from the coverslip.

Of course, sedimentation with hydrodynamic interactions and
Brownian motion is not a protocol followed in simulations of RCP
states.  The algorithm developed by Lubachevsky and Stillinger
considers hard particles moving ballistically \cite{Lubachevsky}.  The particles
start very small, and continue interacting as they gradually
are swelled until the system jams.
The method of O'Hern and co-workers is similar, starting with
small particles that grow, but their particles are not infinitely
hard, nor do they have velocities \cite{OHern}.  Rather, the
simulation proceeds until the particles are maximally swelled but
non-overlapping, thus giving the final hard-sphere state.
Tobochnik and Chapin devised a similar algorithm which used Monte
Carlo moves to eliminate overlaps \cite{Tobochnik}.
These ``expand and eliminate the overlap'' methods
are similar to an earlier method due to Jodrey and Tory which
slowly shrank spheres, sliding pairs of spheres linearly to minimize
their overlap, until all spheres had no overlaps \cite{Jodrey}.
These methods all have the strength that the RCP state is generated
isotropically, in contrast to our experiment where gravity sets
a direction.  (As discussed below, we do not see anything special
about the direction of gravity in our data.)  Our experimental
method does have the feature that our spheres never overlap, in
contrast to algorithms where overlaps are allowed at
intermediate stages \cite{Jodrey,OHern,Tobochnik,hermes10}, although it is not
obvious that intermediate stage overlaps would cause substantially
different results in the final state.  In some ways our
experimental protocol is similar to a method by Visscher and
Bolsterli from 1972 \cite{visscher}.  Their algorithm dropped
particles at random positions until the particles collided with the floor
or a previous particle; the falling particle then rolls downhill
until it reaches a locally stable position.  In our experiment,
all the particles fall
simultaneously, and also their Brownian motion gives them the
ability to find better packings than the Visscher and Bolsterli
algorithm.

\subsection{Connection of 3D images}

\begin{figure}[htbp]
\begin{center}
\includegraphics[width=8cm]{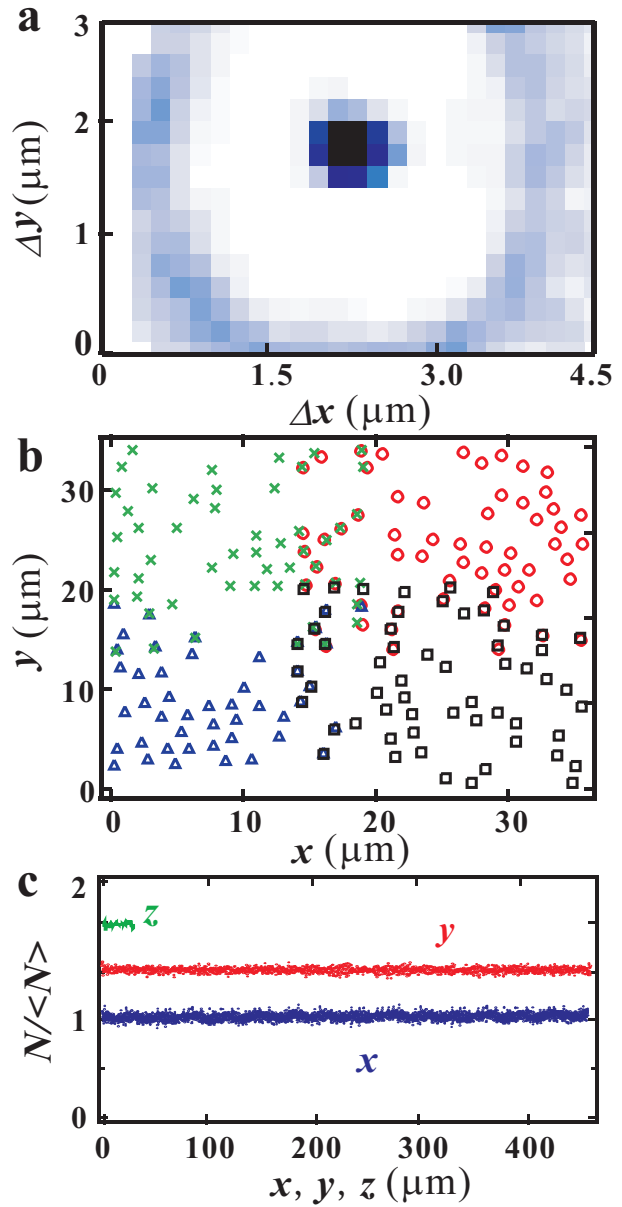}
\end{center}
\caption{(color online) (a) The image plot of the fraction of
successfully superimposed particles $f$ in a plane of ($\Delta x$,
$\Delta y$). The dark central region corresponds to $f$ = 1, meaning that
all possible particle overlaps are successful.  (b) The circles,
triangles, squares and crosses correspond to particle positions
obtained from 4 separate 3D images.  (c) The local number of
particles observed $N$ normalized by the average, as a function
of each axis after connecting the images.  We add an offset to the data of $y$
and $z$ so they can be seen clearly. 
}
\label{connect}
\end{figure}

To take a large ensemble, we scan a grid of 3D images with a
small amount of overlap in $x$ and $y$.  We compute particle
positions from each image and then we connect one image to
an adjacent overlapping image.  Particles are considered as
superimposed when $|\vec{r}_{ij} - \vec{r}_{lk}| <$ 0.2 $\mu$m
where $\vec{r}_{ij}$ is the position of particle $i$ in image $j$
and $\vec{r}_{lk}$ is the position of particle $l$ in adjacent
image $k$.  To achieve this, we apply small displacement shifts
$\Delta x$, $\Delta y$ and $\Delta z$ to one image, and look for the
fraction $f$ of superimposed particles within the overlapped zones.
Figure \ref{connect}(a) shows $f$ in a ($\Delta x$, $\Delta y$)
plane with the resolution of one pixel accuracy, and we find one 
spot where $f \sim 1$. The secondary ring around the central 
spot corresponds to the first peak of the pair correlation function, 
where some coincidences between particle positions are expected.
While Fig.~\ref{connect}(a) shows $f$ in a two-dimensional plane,
we calculate $f$ using shifts in the $z$ direction as well. 
We next apply sub-pixel displacement shifts around the spot 
in Fig.~\ref{connect}(a) to better resolve the peak.
Finally, we calculate the sum of the squared distances
between the positions of the overlapped particles within
each region, and find the global choices of shift values
that minimize the overall squared error, to provide the most
accurate shift factors for the overlap.  Using the shift factors,
we then link up the particle positions in adjacent sections.
The coincident particles are replaced by their average position.
Figure \ref{connect}(b) shows particle positions at $5 < z <
5.2$ $\mu$m after connecting 4 separate overlapping images.
The particle positions are well-superimposed in the overlapping
regions.  
Our sample chamber contains approximately 500,000,000
particles.  
Using the overlapping image method,
we obtain a large 3D data set with size 
492 $\mu$m $\times$ 513 $\mu$m $\times$ 28 $\mu$m, containing more
than 500,000 particles.  Due to artifacts when identifying
particles near the image edges, we clip the data evenly from the
boundaries and our final data set is $V=492$~$\mu$m $\times$ 513 $\mu$m
$\times$ 23.5 $\mu$m, containing $N=453,136$ particles.  
This gives us $\phi_{RCP} = N (\pi d^3/6) / V = 0.646 \pm
0.020$, with the error bars due to the 1\% uncertainty of the
mean particle diameter.  Our value is in agreement with
simulations that considered polydispersity
\cite{Farr,schaertl94}.

We also examine the average number of particles $N$ observed as
a function of $x$, $y$, and $z$.  To do this as a function of
$x$, we count the particles which are located between $x$ and
$x+0.2$ $\mu$m for a sequence of $x$ values; a similar procedure
is used for $N(y)$ and $N(z)$.  The number of particles $N(x)$
as a function of $x$ is fairly flat, as are $N(y)$ and $N(z)$,
as shown in Fig.~\ref{connect}(c).  
However, there are small residual oscillations in $x$ with the
standard deviation of $N(x)/\langle N \rangle$ being 0.027 and
a period of approximately 33~$\mu$m$\approx 13d$.
This is an artifact of our connection algorithm, as we connect
the images along $y$ direction first, then we connect them along
the $x$ direction.  If we change this order, we find $N(x)$ becomes
flat and $N(y)$ undulates.  To evaluate effects of this oscillation,
we calculate the structure factor and the pair correlation function
using both connection orders ($x$ first or $y$ first), and find
almost identical results.  Thus, we ignore these oscillations.

\subsection{Detection of ordered particles} \label{sec:loc}

We use a rotationally invariant local bond order parameters $d_6$
to look for crystalline particles \cite{Steinhardt,Wolde,
Gasser}.  The idea is to calculate for each particle a complex
vector $q_{6m}(i)$, whose components $m$ depend on the
orientation of the neighbors of particle $i$ relative to $i$.
Each of the 13 components of the vector is given by:
\begin{equation}
q_{6m}(i) = \frac{1}{N_b}{\sum^{N_b}_{j=1} Y_{6m}(\hat{r}_{ij})},
\end{equation}
where $N_b$ is the number of nearest neighbor particles for
particle $i$, $\hat{r}_{ij}$ is the unit vector pointing from
particle $i$ to its $j$th neighbor, and $Y_{lm}$ is a spherical
harmonic function.  The $q_{6m}$ parameters are the
coefficients for the spherical harmonics in an expansion of the
vector directions $\hat{r}_{ij}$, and thus capture a sense of the
structure around particle $i$.  The $l=6$ harmonics are used as
it is known that on a local level, hexagonal symmetry is often
present due to packing constraints \cite{Steinhardt,Wolde}.
The neighbors of a particle are defined as those with centers
separated by less than $1.4d$ (which is the location of the
first minimum of the pair correlation function).
These 13-dimensional complex vectors are then
normalized to unity using
\begin{equation}
\hat{q}_{6m}(i) = \frac{q_{6m}(i)}{(\sum_{m}q_{6m}(i)\cdot
q^*_{6m}(i) )^{0.5}}.
\end{equation}
Then, we compute $d_6$ as:
\begin{equation}
d_6(i,j) = \sum_{m=-6}^6q_{6m}(i)\cdot q^*_{6m}(j).
\end{equation}
$d_6(i,j)$ is a normalized quantity correlating the local
environments of neighboring $i$ and $j$ particles. $d_6(i,j)$ is a
scalar and its range is $-1 \le d_6(i,j) \le 1$;
$d_6(i,j)=1$
would correspond to two particles who have identical local
environments, at least identical in the sense captured by the
$q_{6m}$ data.  Two neighboring
particles are termed ``ordered neighbors" if $d_6(i,j) > 0.5$.
The number of ordered neighbors $N_o^i$ is decided for each
particle.  $N_o^i$ measures the amount of similarity of structure
around neighboring particles.  $N_o^i$=0 corresponds to random
structure around particle $i$, while a large value of $N_o^i$
means that particle $i$ and its neighbor particles have similar
surroundings.  Following prior work, particles with $N_o^i \ge 8$
are classed as crystalline particles, and the other particles are
liquid-like particles \cite{Wolde}.

We also compute the $\hat{W}_l^i$ parameter to specify local structures:
face centered cubic (fcc),
icosahedral structure (icos), hexagonal close packed (hcp)
and body centered cubic (bcc) \cite{Steinhardt}.
The $\hat{W}_l^i$ parameter is defined as
\begin{eqnarray}
\bar{Q}^i_{lm} &\equiv& \langle Y_{lm}(\hat{r}_{ij}) \rangle \\
W_l^i &=& \sum_{m_1,m_2,m_3} \left (
\begin{array}{ccc}
l & l & l \\
m_1 & m_2 & m_3
\end{array}
\right ) \bar{Q}_{lm_1}^i\bar{Q}_{lm_2}^i\bar{Q}_{lm_3}^i
\end{eqnarray}
where $\langle \rangle$ corresponds to the average over neighboring
particles $j$, $m_1 + m_2 + m_3 = 0$, and
\begin{eqnarray}
\hat{W}_l^i &\equiv& W_l^i / \left ( \sum^l_{m=-l} |\bar{Q}^i_{lm}|^2 \right )^{3/2}.
\end{eqnarray}
The coefficients
\begin{eqnarray}
\left (
\begin{array}{ccc}
l & l & l \\
m_1 & m_2 & m_3
\end{array}
\right ) \nonumber
\end{eqnarray}
are Wigner $3j$ symbols.  Similar to the $q_{6m}$ parameters
discussed above, the $\hat{W}_l$ parameters are able to capture a
sense of the local ordering with $l$-fold symmetry, and have been
used before to help classify local structure; see
Ref.~\cite{Steinhardt}.
The values of $\hat{W}_l^i$ for ideal
structures are listed in Table \ref{Wl} \cite{Steinhardt}.
These ideal structures are unrealistic for experimental data,
so we generate 50,000 representations of each ordered structure
and perturb their positions by 5 \% of the particle diameter,
to match the polydispersity of our experimental particle sizes.
This gives us a distribution of $\hat{W}_l^i$ for each ordered
structure (Table \ref{Wl2}).  Within our experimental data,
we calculate $\hat{W}_l^i$ ($l$ = 4, 6, 8) for each particle.
A particle is classed as a ordered particle when $\hat{W}_4$,
$\hat{W}_6$ and $\hat{W}_8$ of a particle are simultaneously within
the ranges of one structure shown in Table \ref{Wl2}.  Otherwise,
particles are classed as random particles.

\begin{table}
\begin{center}
\caption{The values of $\hat{W_l}$ ($l$ =4, 6, 8) for
ideal structures of fcc, icosahedron, hcp and bcc \cite{Steinhardt}.
We add the notation of (i) for each ideal structure.
}
\begin{ruledtabular}
\begin{tabular}{lccc}
& $\hat{W}_4$ & $\hat{W}_6$ & $\hat{W}_8$ \\
\hline
fcc(i) &  -0.159316 & -0.013161 & 0.058454 \\
icos(i) & & -0.169754 &  \\
hcp(i) & 0.134097 & -0.012442 & 0.051259 \\
bcc(i) & 0.159317 & 0.013161 & -0.058455 \\
\end{tabular}
\end{ruledtabular}
\label{Wl}
\end{center}
\end{table}

\begin{table}
\begin{center}
\caption{The ranges of values of $\hat{W_l}$ ($l$ =4, 6, 8) for
structures with 5\% perturbations
from ideal structures \cite{Steinhardt}.
We add the notation of (p) for the perturbed structures.
}
\begin{ruledtabular}
\begin{tabular}{lccc}
& $\hat{W}_4$ & $\hat{W}_6$ & $\hat{W}_8$ \\
\hline
fcc(p) & -0.085 $\sim$ -0.169 & -0.0109 $\sim$ -0.0193 & -0.0180 $\sim$ 0.0640 \\
icos(p) & -0.050 $\sim$ 0.200 & -0.171 $\sim$ -0.162 & -0.090 $\sim$ 0.090 \\
hcp(p) & 0.067 $\sim$ 0.138 & -0.036 $\sim$ -0.004 & 0.000 $\sim$ 0.080 \\
bcc(p) & 0.152 $\sim$ 0.161 & -0.015 $\sim$ 0.021 & -0.072 $\sim$ 0.060 \\
\end{tabular}
\end{ruledtabular}
\label{Wl2}
\end{center}
\end{table}

\subsection{Calculating the structure factor}
\label{sec:sq}

We compute the structure factor $S(\vec{q})$ via a direct Fourier
transform of the particle position, $S(\vec{q}) = 
N^{-1}|\sum^N_{i=1} \exp(i \vec{q} \cdot \vec{r}_i)|^2$
where $\vec{r}_i$ is the particle position.  $S(q)$ is the average
of $S(\vec{q})$ over $q = \vec{q}$.

Our large images have two advantages for calculating the structure
factor $S(q)$.  The first is a high resolution with respect to $q$,
as the resolution is given by $\delta q = 2 \pi / L$ where $L$ is
the image size.  Our sample size is 492 $\mu$m $\times$ 513 $\mu$m
$\times$ 23.5 $\mu$m and this yields $\delta q = 0.0128$~$\mu m^{-1}$.
The second advantage of a large image is the reduction of boundary
effects.  Our experimental data set does not obey periodic boundary
conditions, unlike most simulations.  Thus, we need to use a
window function to minimize the influence of the data cutting off
at the boundaries, or we need to periodically replicate the data.
Both these procedures increase $S(q)$ only near $q=0$, but it is
precisely $S(q=0)$ that is of interest.  
Larger images allow us to
go to smaller $q$ with less problems.
We checked the Hann window,
Hamming window, the Blackman window, and also using no window function.
We find that $S(q)$ varies for $q < 0.55$~$\mu m^{-1}$, corresponding to $qd/2\pi
= 0.2$.  That is, our results for $q > 0.55$~$\mu m^{-1}$ are
independent of our choice of window functions.  In what follows, we
do not use a window function, and will focus on the results for
small $q$ but considering only $q > 0.55$~$\mu m^{-1}$.

\section{Results}
\subsection{Minimal local ordering}

\begin{figure}
\begin{center}
\includegraphics[width=8cm]{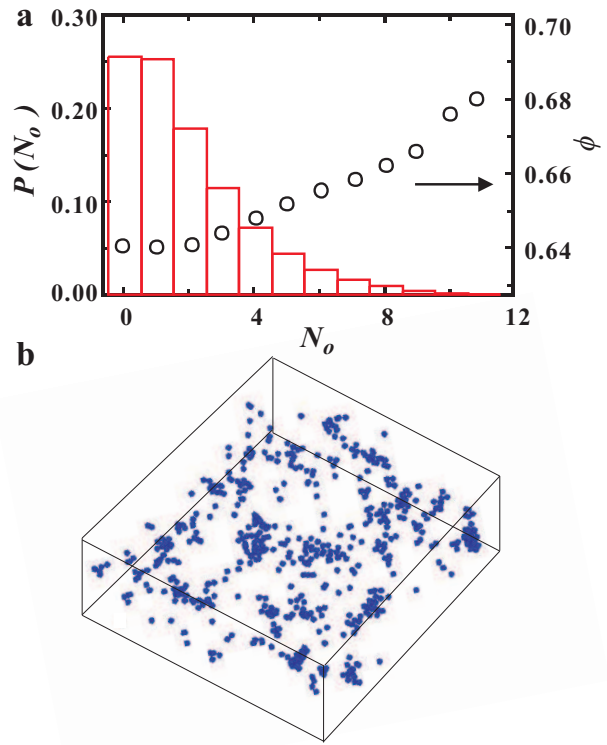}
\end{center}
\caption{(color online). (a) The probability of the number of
ordered neighbors $N_o$.  When a particle has $N_o^i \ge 8$, it
is classed as crystalline. The circles corresponds to local volume
fraction calculated from Voronoi cell volume, averaged over
all particles with $N_o$ having the given value. 
(b) The spatial distribution of crystalline
particles in 3D image.  This image is 115 $\mu$m $\times$ 115
$\mu$m $\times$ 23.5 $\mu$m.  The other particles are not drawn,
to better show the crystalline particles.
}
\label{local}
\end{figure}

First, we investigate the randomness of our sample. 
We compute the fraction of ordered neighbors in the sediment 
of our colloidal suspension using the $d_6$ parameter described in
Sec.~\ref{sec:loc}. 
Fig.~\ref{local}(a) shows the probability of finding particles
with a given number of ordered neighbors $N_o$.
Following prior works, particles with $N_o^i \ge 8$ are classed
as crystalline particles, and the other particles are liquid-like
particles \cite{Wolde}.  We find the fraction of crystalline
particles is below 0.03, and that these
particles are well-dispersed throughout the sample,
and shown in Fig.~\ref{local}(b).  At most, we see small
crystallites composed of 3 or 4 crystalline particles which
are nearest neighbors.  Furthermore,
the fraction of particles which $N_o$ is below 3 is over 0.8.
It means that the coordinate particle arrangement of over 80\% of
particles are not similar to those of nearest neighbor particles.
The effects on the structure by the spatial distribution of crystalline particles 
will be discussed below.
We consider that our system is essentially randomly packed 
as the crystalline particles are a quite low fraction and well-dispersed. 

We also compute the fraction of specific local ordered structures:
fcc, icosahedron structure (icos), hcp and bcc.  The importance
of those structures was emphasized over 50 years ago by Frank
\cite{Frank}; for example, the icosahedral arrangement has a
significantly lower energy than an hcp or fcc cluster for simple
Lennard-Jones potentials.  To specify local ordered structure,
we compute $\hat{W}_l^i$ (l = 4, 6, 8) parameters for each
particle \cite{Steinhardt} (see Sec.~\ref{sec:loc}).  We find
that the fraction of particles that are fcc, icosahedron, hcp
and bcc are 0.0020, 0.0001, 0.0066 and 0.0014, respectively.
The sum of those fraction is $\sim$ 0.01 and this is consistent
with the result of $N_o^i$ analysis.  Again, this suggests that
the sample is randomly packed.  In addition, it is interesting
that icosahedron is the least fraction we observe in our packed
hard sphere-like particles, whereas icosahedral structure is most
stable local structure for Lennard-Jones potentials \cite{Frank}.
This is consistent with many prior observations, and recent
simulations suggest that icosahedral structures are indeed not as
relevant for random close packed spheres as polytetrahedrons are
more favored local structures \cite{vanmeel09}.  We also find that
the fraction of hcp ordering in our sample is higher than that of fcc.

\subsection{Voronoi cell volume distribution}

Next, we study the local volume fraction of the sediment 
at the particle length scale. 
We compute the Voronoi decomposition which is
a unique partitioning of space.  Each particle is within its
own Voronoi polyhedron, and the Voronoi polyhedron is the region
of space which is closer to the given particle than any other particle
\cite{Preparata}.
We calculate a volume for each Voronoi cell except for
those cells located
on the boundaries, which have incorrectly defined volumes.

We compute the local volume fraction for each particle as 
$\phi_i = \pi \langle d \rangle^3 /6 V_i$ 
where $\phi_i$ and $V_i$ are the local volume fraction and 
Voronoi cell volume for particle $i$, respectively. 
We use the average diameter $d$
since we can not detect each particle diameter.
The circles in Fig.~\ref{local}(a) show the average local volume fraction 
as a function of $N_o$. 
We find that the local volume fraction is almost constant at $N_o \le$ 2, 
but increases with larger $N_o$. 
This result means that few highly ordered particles have a higher local 
volume fraction than random particles. 
It is natural since ordered phase such as fcc crystal is the most packed phase 
and this tendency is suggested by previous reports \cite{Perera, Luchnikov, Eric2002}.

\begin{figure}[ht]
\begin{center}
\includegraphics[width=8cm]{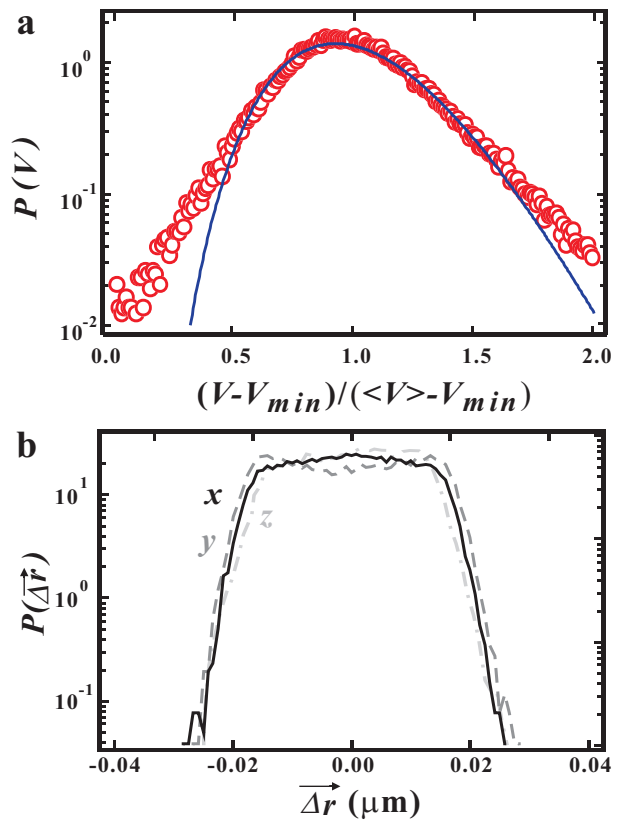}
\end{center}
\caption{(color online) (a) Distribution of the Voronoi cell volumes plotted
as a function of
$(V-V_{min})/(\langle V \rangle - V_{min})$.  The solid
line is a fitting line with Eq.~\ref{voleq} using
$k=13.1$.
(b) Distribution of the position differences
between the particle position and the center of mass of the Voronoi cell.
Almost all particles are located at the center of mass of the Voronoi cell.
The values on the horizontal axis only go from -0.04 to +0.04 $\mu$m, 
much less than the particle diameter $d=2.53$~$\mu$m.
}
\label{voronoi}
\end{figure}

Next, we compute a distribution of Voronoi cell volume. 
Aste and coworkers proposed a universal function of the
distribution of Voronoi cell volumes \cite{Aste}, and the form is 
described as 
\begin{eqnarray}
P(V,k) = \frac{k^k}{(k-1)!}\frac{(V-V_{min})^{k-1}}{(\langle V \rangle - V_{min})^k}
\exp \left (-k\frac{V-V_{min}}{\langle V \rangle - V_{min}} \right )
\label{voleq}
\end{eqnarray}
where $\langle V \rangle$ is the average of the Voronoi cell
volumes.  It is worth noting that the only adjustable parameter in
Eq.~\ref{voleq} is $k$, other than $V_{min}$ which is constrained.
$k$ is termed the ``shape parameter'' and corresponds the
number of elementary cells composing the Voronoi cell \cite{Aste}.
For instance, the value of $k$ is 1 in an fcc crystal, while $k$ is
close to the number of nearest neighbor particles (about 12 or 13)
in random structure \cite{Aste}.  We choose $V_{min} = 0.694 d^3$,
which is the smallest Voronoi cell that can be built in a packing
of monodisperse spheres \cite{Aste}.  Figure \ref{voronoi}(a)
shows a distribution of the Voronoi cell volumes as a function
of $(V-V_{min})/(\langle V \rangle - V_{min})$.  The shape of
the distribution is asymmetric and not Gaussian, that is, the
distribution is narrow at small volumes and broad at large volumes.
We fit the distribution of Voronoi cell volumes with Eq.~\ref{voleq}
and obtain $k$ = 13.1 (the solid line in Fig.~\ref{voronoi}).
The tail of the distribution is broader than the fitting line,
perhaps due to the 5\% polydispersity of our particles.  The $k$
value was investigated in experiments using small glass beads
($\sim$ 250 $\mu$m) in water \cite{Aste}, acrylic spheres with
different preparation methods \cite{Aste} and larger glass beads
($\sim$ 3 mm) in oil \cite{Losert}.  Those similar experiments
found 11 $\le k \le$ 13 \cite{Aste} and close to $k$ = 14.2 $\pm$
0.6 \cite{Losert} for random sphere packing.  $k$ varies with
each experiment since $k$ slightly depends on the polydispersity.
Our experimentally observed value of $k$ = 13.1 is consistent
with those prior experiments.  This is further evidence that the
arrangement of our sample is random. We note that a universality
of $k$ value is proposed of the distribution of Voronoi cell
volumes for random sphere packing, with the evidence coming from
experimental with non Brownian particles \cite{Aste, Losert}.
Our agreement with the prior work suggests that our close-packed
sediment is not strongly different despite the Brownian motion
that the particles have during sedimentation.

Within each Voronoi cell, we now consider the positions of particles
relative to the Voronoi cell ``center of mass.''
We compute a vector $\vec{\Delta r}_i \equiv
\vec{r}_i - \vec{g}_i$ where $\vec{r}_i$ is the position vector
for particle $i$ and $\vec{g}_i$ is the position vector for
the center of mass of the Voronoi cell which include particle $i$.
Figure \ref{voronoi}(b) shows the distribution of each axis component
of $\vec{\Delta r}_i$.  We find almost all particles are located at the
centers of their Voronoi cells within the resolution of particle tracking
($\sim$ 0.05 $\mu$m), even along the direction of gravity.

\subsection{Density fluctuations}

\begin{figure}
\begin{center}
\includegraphics[width=8cm]{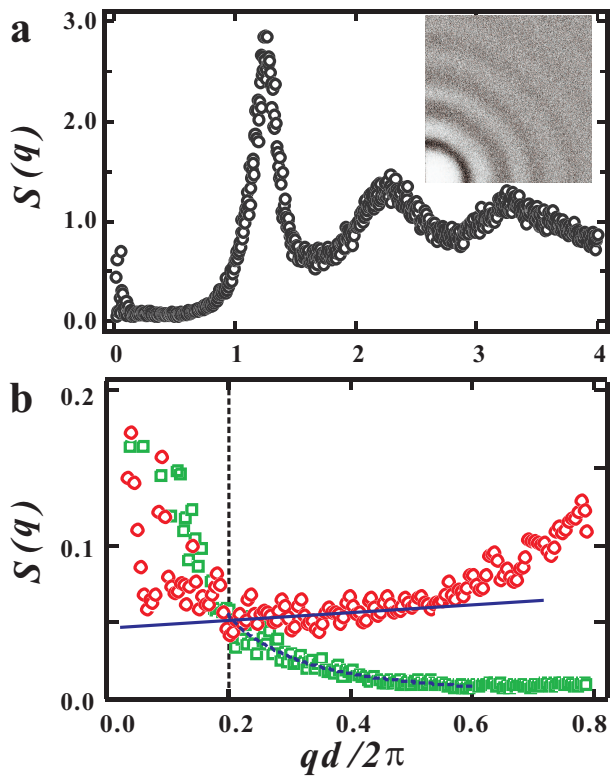}
\end{center}
\caption{(color online) 
(a) The structure factor $S(q)$ as a function of wavenumber $q$.
The inset is the quarter image of the structure factor in a plane of
($q_x$, $q_y$).  (b) The expanded view of $S(q)$ (circles)
and $S_c(q)$ (squares) near $q$ = 0. 
$S(q)$ increases near $q$ = 0 because of computational
artifacts due to the finite size of our data set; 
the data are reliable for $qd/2\pi > 0.2$, indicated by
the vertical dashed line.  The solid line is a fitting
line with a linear function over the data with $0.2 < qd/2\pi < 0.5$.  We obtain $S(0)$ =
0.049 by interpolating the fitting line to $q$ = 0.
On the other hand, the
structure factor for the crystalline particles $S_c(q)$ decreases with larger $q$ and 
it is well fitted by Ornstein-Zernike function (dashed line) over $0.2 < qd/2\pi < 0.6$. 
}
\label{sq}
\end{figure}

We next check whether the sediment is
in a ``strictly
jammed state" or not.  As we mentioned above, $S(0)$ = 0 is 
required in strict jamming states since 
strict jamming states should be incompressible 
(equivalently,
hyperuniform \cite{Torquato2005}). 
To obtain $S(0)$ value, we directly calculate $S(\vec{q})$ from the
particle positions.  The inset in Fig.~\ref{sq}(a) shows a image
plot of the structure factor in a plane of ($q_x$, $q_y$) where $q_x$
and $q_y$ are the $x$ and $y$ components of vector $q$.  $S(\vec{q})$
is quite isotropic even though $y$ is the direction of gravity,
again further implying our sediment is randomly packed.
We average $S(\vec{q})$ over $q = |\vec{q}|$ and obtain $S(q)$,
shown in Fig.~\ref{sq}(a).  Figure \ref{sq}(b) shows $S(q)$
near $q$=0 (circles).  $S(q)$ increases near $q$ = 0 because of
computational artifacts (see Sec.~\ref{sec:sq}); we find
that $S(q)$ is reliable over $qd/2\pi \ge 0.2$, indicated
by the vertical dashed line in the figure.  We fit $S(q)$
with a linear function between $0.2 \le qd/2\pi \le 0.5$ and obtain
$S(0) = 0.049 \pm 0.008$ by extrapolation.  $S(q)$ is also well fitted by a
parabolic function between $0.2 \le qd/2\pi \le 0.5$ and $S(0)$ is
almost the same.  Our data are
insufficiently strong to determine if the linear fit or
parabolic fit is more reasonable \cite{Torquato2005}.  Our uncertainty (0.008) is
determined by trying the different Fourier transform windowing functions, in
combination with linear or parabolic fits:  all possible combinations yield
values within the range $S(0) = 0.049 \pm 0.008$, and thus we state with confidence
that $S(0) \ne 0$.
Donev, Stillinger and Torquato obtained
$S(0)$ = 6.1 $\times$ $10^{-4}$ by numerical simulation with one
million monodisperse particles \cite{Torquato2005} and 
our experimental value of $S(0)$ is about 100 times larger than simulation result,
a significant difference well beyond the uncertainty of our data.
This implies that our sample is much more compressive than the
structure found by simulation.

\begin{figure}
\begin{center}
\includegraphics[width=8cm]{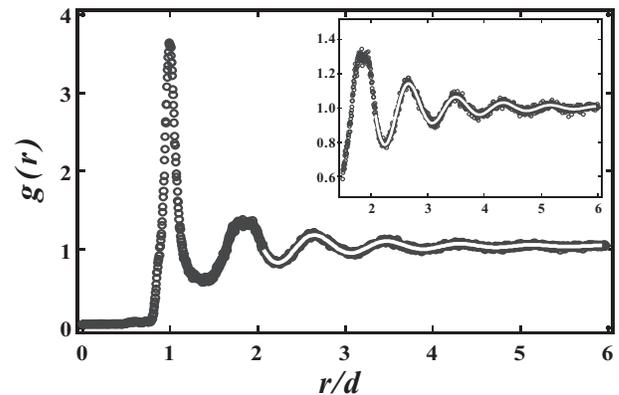}
\end{center}
\caption{The pair correlation function $g(r)$ of the sediment.
The inset is enlarged view at 2 $< r/d < $ 6.
The solid line is a fitting line with Eq.~\ref{gre} over $r/d > 2$.
}
\label{gr}
\end{figure}

To support this result, we use a real space function: the pair
correlation function $g(r)$ shown in Fig.~\ref{gr}.  $g(r)$ at
$r/d >$ 2 is well fitted by a exponentially damped oscillatory
function \cite{Perry, Torquato2002} described as: \begin{eqnarray}
g(r) \sim \frac{C}{r}\exp(-r/\xi)\cos[K_0(r-r_0)] +1 \label{gre}
\end{eqnarray} where $C$, $\xi$ and $K_0$ correspond to an
amplitude, a characteristic length of spatial correlation and
the period of the oscillations, respectively.  From the fitting,
we obtain $C$ = 2.27 $\pm$ 0.08, $\xi = 1.50d \pm 0.03d$
and $K_0 = 7.55/d \pm 0.01/d$.  Again, we compare with the
simulation of a strictly jammed state which yields $\xi = 1.83d$
and $K_0 = 7.58/d$ \cite{Torquato2005}.  Though $K_0$ is similar
between experiment and simulation, the length scale $\xi$ from
our experiment is 
shorter than that of simulation.  The decay of $g(r)$ in an experiment
is related to the broadness of each peak, that is, $g(r)$ decays
quickly when peaks are slightly broad.  Broad peaks mean that
a distance between two particles are distributed.  Hence, the
rapid decay of $g(r)$ are also connected with the fluctuations in 
density, supporting the result
$S(0) \neq 0$.  Thus, we conclude that the
arrangement in our experiment is not a strictly jammed state.
It is important to note that uncertainty in
particle positions will broaden the first peak of $g(r)$, but
this does not strongly affect $g(r)$ for larger $r$ as those
uncertainties do not accumulate over large distances.  That is,
the true separation between two particles has a distance $r_{ij}$
with an uncertainty of $\pm 0.06$~$\mu$m, which is somewhat
significant when $r_{ij}$ is small and less important when
$r_{ij}$ is large.

There are several possible explanations for this observed
``softness.''  One possibility is the polydispersity of colloidal
size which is a crucial reality for experimental situations, and a
difference with the simulations to which we are comparing our data.
When particle sizes are slightly different, the minimum distance
between two particles can be changed from $\langle d \rangle$ and
then the first peak of $g(r)$ becomes slightly broad, consistent
with Fig.~\ref{gr}.  This small difference adds up over a long
distance and then it may induce long wavelength fluctuations.  In
particular, the local fluctuations in composition (slightly more
large particles or small particles) are coupled to the number
fluctuations.  
A polydispersity of 5\% such as we have in our
experiment results in $S(q\rightarrow 0) = 0.04 \pm 0.01$
based on simulations \cite{ludovic}, consistent with our data.
Unfortunately, we cannot determine the individual particle sizes
as the resolution of optical microscopy blurs particle images on
the same scale as the variability of particle size.  Furthermore,
images of neighboring particles overlap, again due to the finite optical
resolution.  This makes determining individual particle size
problematic, and prevents us from disentangling the influence of
polydispersity from our data.

A second possibility is that our sample is not at random close
packing due to friction effects between the particles, which is
quite important for granular packings.  It is known that granular
packings are often looser than RCP, with volume fractions as
low as $\phi \approx 0.58$, termed ``random loose packing''
\cite{Onoda}.  By vibrating the system, the packing fraction can
be increased, perhaps coming close to RCP \cite{Bernal,knight95}.
In our experiment, particles move by Brownian motion, and this
may let them find the RCP state.  Furthermore, the particles are
sterically stabilized to prevent them from sticking together.
In general, friction is not a concept that is usually applied to
colloidal particles.  However, we cannot completely rule out the
possibility of some possible occasional attractive interaction
between our particles which might result in friction-like behavior,
resulting in a slightly loose packing.  Small amounts of static
friction gave nonzero $S(0)$ values in simulations \cite{Silbert}.

A third possibility is based on the $N_o$ dependence of the local
volume fraction (Fig.~\ref{local}(a)), that is, particles in more
ordered local environments are packed better.  The crystalline
particles, which have high local volume fraction, are distributed
throughout the sample (see Fig.~\ref{local}(b)).  To quantify the
spatial distribution of the crystalline particles, we calculate
a crystalline structure factor $S_c(q)$ as $S_c(\vec{q}) =
{N}^{-1}|\sum^{N}_{i=1} W_i \exp(i \vec{q} \cdot \vec{r}_i)|^2$
where $W_i$ = 1 when $i$ particle is classed as crystalline,
otherwise $W_i$ = 0.  $S_c(q)$ is the average of $S_c(\vec{q})$
over $q = \vec{q}$.  The square symbols in Fig.~\ref{sq}(b)
correspond to $S_c(q)$ and we find that $S_c(q)$ can be fitted with
Ornstein-Zernike function [dashed line in Fig.~\ref{sq}(b)].  This
fit gives us that the typical length scale between the crystalline
particles is $12.8d \pm 1.8d$.  Thus, the spatial distribution of
crystalline particles can induce density fluctuations with long
wavelength and it might be another reason for our observation that
$S(0) \neq 0$.  It is worth noting that the small but nonzero
fraction of the crystallites (less than 3\%) is crucial in this
conjecture.  It is possible that these tiny crystallites are due
to Brownian motion during the sedimentation.  We are unaware of
any measurements of tiny crystallites in simulations of random
close packing, although one recent study of a binary mixture of
spheres used the same order parameter that we do and found the
average number of ordered bonds (our $N_o$) was small \cite{Ken}.

\section{Conclusions}

We use confocal microscopy to study both the local and long-range
structure of a random close packed colloidal suspension.  We find
that the fraction of crystalline particles is at most 3\%, and
furthermore that almost no regions in the sample have icosohedral
order (less than 0.01\%).  These observations suggest that the
sample is randomly packed.  This is further supported by the
distribution of Voronoi volumes, which is well fit by a prediction
based on a model of random packing.

We also compute the static structure factor $S(q)$ and find that
$S(0) = 0.05$, in contrast to simulations which found $S(0) =
6 \times 10^{-4}$ \cite{Torquato2005}.  $S(0)$ is proportional to
the isothermal compressibility, implying that the simulated states
are essentially incompressible (to within numerical precision),
while our experimental sample is compressible.  This may be due
to the presence of tiny crystalline regions in our sample, which
are associated with slightly higher local density (and thus give
rise to long wavelength density fluctuations).  Alternatively,
it may be due to the polydispersity of particle sizes in the
experiment ($\sim 5$\%).  This softness ($S(0) \neq 0$) is
crucial to how the sample would respond to an external force,
for example, shear stress.  The viscosity and elasticity of
a sample are extremely sensitive to density near jamming point
\cite{Olsson, OHern}.  Near the jammed state, small fluctuations in
density result in large fluctuations of viscosity and elasticity,
which can lead to shear instability or cracking \cite{Furukawa}.
This suggests that real-world RCP materials may possess nontrivially
different properties from idealized simulations.  Our work points
to polydispersity and sample preparation as the possible origin of
these differences, both of which are worthy of further exploration
in both simulation and experiment.

\section*{Acknowledgments}
We thank L.~Berthier and P.~Charbonneau for helpful discussions,
and we thank G.~Cianci and K.~Edmond for making our colloids.
R.~K.~was supported by a JSPS Postdoctoral Fellowship for Research Abroad.
E.~R.~W.~was supported by a grant from the National Science
Foundation (DMR-0804174).


\begin{thebibliography}{99}
\bibitem{Bernal} J. D. Bernal and J. Mason, Nature \textbf{188,} 910 (1960).
\bibitem{Finney} J. L. Finney, Proc. R. Soc. London A {\bf 319,} 479 (1970). 
\bibitem{Scott} G. D. Scott and D. M. Kilgour, 
Br. J. Appl. Phys. \textbf{2,} 863 (1969).
\bibitem{Jodrey} W. S. Jodrey and E. M. Tory, Phys. Rev. A 
\textbf{32,} 2347 (1985).
\bibitem{Tobochnik} J. Tobochnik and P. M. Chapin, 
J. Chem. Phys. \textbf{88,} 5824 (1988).
\bibitem{Lubachevsky} B. D. Lubachevsky and F. H. Stillinger, 
J. Stat. Phys. \textbf{60,} 561 (1990).
\bibitem{Torquato2005} A. Donev, F. H. Stillinger and S. Torquato,
Phys. Rev. Lett. \textbf{95,} 090604 (2005).
\bibitem{chaudhuri10} P. Chaudhuri, L. Berthier, and S. Sastry,
Phys. Rev. Lett. \textbf{104}, 165701 (2010)
\bibitem{radin08} C. Radin, J. Stat. Phys. \textbf{131}, 567 (2008).
\bibitem{Krieger} I. M. Krieger, Adv. Colloid Interface Sci. 
{\bf 3,} 111 (1972). 
\bibitem{Torquato2000} S. Torquato, T. M. Truskett and P. G. Debenedetti, 
Phys. Rev. Lett \textbf{84,} 2064 (2000).
\bibitem{Hales} J. C. Hales, Ann. Math. \textbf{162,} 1065 (2005).
\bibitem{Davis} K. E. Davis, W. B. Russel and W. J. Glantschnig, Science \textbf{245,} 507 (1989).
\bibitem{Pouliquen} O. Pouliquen, M. Nicolas and P. D. Weidman, Phys. Rev. Lett \textbf{79,} 3640 (1997).
\bibitem{Torquato2003} S. Torquato and F. H. Stillinger, Phys. Rev. E \textbf{68,} 041113 (2003).
\textbf{68,} 069901 (2003).
\bibitem{Cates} M. E. Cates, J. P. Wittmer, J.-P. Bouchaud and P. Claudin, 
Phys. Rev. Lett. \textbf{81,} 1841 (1998).
\bibitem{Liu} A. J. Liu and S. R. Nagel, Nature (London) \textbf{396,} 6706 (1998). 
\bibitem{Silbert} L. E. Silbert and M. Silbert, Phys. Rev. E {\bf 80,} 041304 (2009). 
\bibitem{smith29} W. O. Smith, P. D. Foote, and P. F. Busang,
Phys. Rev. \textbf{34}, 1271 (1929).
\bibitem{bernal64} J. D. Bernal, Proc. Roy. Soc.
London. Ser. A \textbf{280}, 299 (1964).
\bibitem{Losert} S. Slotterback, M. Toiya, L. Goff, J. F. Douglas and 
W. Losert, Phys. Rev. Lett.  \textbf{101,} 258001 (2008).
\bibitem{Aste} T. Aste, T. D. Matteo, M. Saadatfar, T. J. Senden,
M. Schr{\"o}ter and H. L. Swinney, Euro. Phys. Lett. \textbf{79,} 24003 (2007).
\bibitem{epaps} See supplementary material at [URL will be
inserted by AIP] for a file of the particle coordinates.
\bibitem{Alder} B. J. Alder and T. E. Wainwright, J. Chem. Phys. \textbf{27,} 1208 (1957).
\bibitem{Pusey} P. N. Pusey and W. van Magen, Nature (London) \textbf{320,} 340 (1986).
\bibitem{Antl}  L. Antl, J. W. Goodwin, R. D. Hill, R. H. Ottewill, S. M.
Owens, S. Papworth, and J. A. Waters, Colloid Surf. \textbf{17,} 67 (1986).
\bibitem{Dinsmore}A. D. Dinsmore, E. R. Weeks, V. Prasad, A. C. Levitt,
and D. A. Weitz, Appl. Opt. \textbf{40,} 4152 (2001). 
\bibitem{Farr} R. S. Farr and R. D. Groot, J. Phys. Chem. {\bf 131,} 244104 (2009).
\bibitem{schaertl94} W. Schaertl and H. Sillescu, J.
Stat. Phys. \textbf{77}, 1007 (1994).
\bibitem{hermes10} M. Hermes and M. Dijkstra, Europhys. Lett.
\textbf{89}, 38005 (2010).
\bibitem{Crocker96} J. C. Crocker and D. G. Grier, J. Colloid Interface Sci. {\bf 179,} 298 (1996).
\bibitem{Segre} P. N. Segr\`e , E. Herbolzheimer and P. M. Chaikin,
Phys. Rev. Lett., {\bf 79,} 2574 (1997).
\bibitem{Wysocki} A. Wysocki, C. P. Royall, R. G. Winkler, G. Gompper, H. Tanaka, A. van Blaaderen and H. L\"{o}wen,
Soft Matter, \textbf{5,} 1340 (2009). 
\bibitem{pusey09} P. N. Pusey, E. Zaccarelli, C. Valeriani, E.
Sanz, W. C. K. Poon, and M. E. Cates, Phil. Trans. Roy. Soc. A
\textbf{367}, 4993 (2009).
\bibitem{fasolo04} M. Fasolo and P. Sollich, Phys. Rev. E
\textbf{70}, 041410 (2004).
\bibitem{auer01} S. Auer and D. Frenkel, Nature \textbf{413}, 711 (2001).
\bibitem{schope07} H. J. Schope, G. Bryant, and W. van Megen, J.
Chem. Phys. \textbf{127}, 084505 (2007).
\bibitem{Paddy} C. P. Royall, J. Dzubiella, M. Schmidt and A. van Blaaderen, Phys. Rev. Lett. \textbf{98,} 188304 (2007).
\bibitem{Marconi} U. Marini Bettolo Marconi and P. Tarazona, J. Chem. Phys. \textbf{110,} 8032 (1999).
\bibitem{Ken} K. W. Desmond and E. R. Weeks, Phys. Rev. E
\textbf{80}, 051305 (2009).
\bibitem{nugent07prl} C. R. Nugent, K. V. Edmond, H. N. Patel and 
E. R. Weeks, Phys. Rev. Lett. {\bf 99,} 025702 (2007). 
\bibitem{OHern} C. S. O'Hern, L. E. Silbert, A. J. Liu and S. R. Nagel
Phys. Rev. E, {\bf 68,} 011306 (2003). 
\bibitem{visscher} W. M. Visscher and M. Bolsterli, Nature {\bf
239}, 504 (1972).
\bibitem{Steinhardt} P. J. Steinhardt, D. R. Nelson and M. Ronchetti, 
Phys. Rev. B \textbf{28,} 784 (1983).
\bibitem{Wolde} P. R. ten Wolde, M. J. Ruiz-Montero and D. Frenkel, 
J. Chem. Phys. \textbf{104,} 9932 (1996).
\bibitem{Gasser} U. Gasser, E. R. Weeks, A. Schofield, P. N. Pusey and 
D. A. Weitz, Science \textbf{292,} 258 (2001).
\bibitem{Frank} F. C. Frank, Proc. R. Soc. London Ser. A
\textbf{215,} 43 (1952).
\bibitem{vanmeel09} J. A. van Meel, D. Frenkel, and P.
Charbonneau, Phys. Rev. E \textbf{79}, 030201(R) (2009).
\bibitem{Preparata} F. P. Preparata and M. I. Shamos, {\it Computational Geometry} (Springer-Verlag, New York, 1985). 
\bibitem{Perera} D. N. Perera and P. Harrowell, J. Chem. Phys. {\bf 111,} 5441 (1999).
\bibitem{Luchnikov} V. A. Luchnikov, N. N. Medvedev, Yu. I. Naberukhin,
V. N. Novikov, Phys. Rev. B {\bf 51,} 15569 (1995). 
\bibitem{Eric2002} E. R. Weeks and D. A. Weitz, Phys. Rev. Lett. \textbf{89,} 095704 (2002).
\bibitem{Perry} P. Perry and G. J. Throop, J. Chem. Phys. \textbf{57,} 1827 (1972).
\bibitem{Torquato2002} S. Torquato and F. H. Stillinger, J. Phys. Chem. B \textbf{106,} 8354 (2002);
\textbf{106,} 11406 (2002). 
\bibitem{ludovic} L.~Berthier, personal communication.
\bibitem{Onoda} G. Y. Onoda and E. G. Liniger, 
Phys. Rev. Lett. \textbf{64,} 2727 (1990).
\bibitem{knight95} J. B. Knight, C. G. Fandrich, C. N. Lau, H. M.
Jaeger, and S. R. Nagel, Phys. Rev. E {\bf 51}, 3957 (1995).
\bibitem{Olsson} P. Olsson and S. Teitel, Phys. Rev. Lett., {\bf 99,} 178001 (2007).
\bibitem{Furukawa} A. Furukawa and H. Tanaka, Nat. Mat. \textbf{8,} 601 (2009).
\end{thebibliography}
\end{document}